\def\year{2021}\relax
\documentclass[letterpaper]{article} % DO NOT CHANGE THIS
\usepackage{aaai21}  % DO NOT CHANGE THIS
\usepackage{times}  % DO NOT CHANGE THIS
\usepackage{helvet} % DO NOT CHANGE THIS
\usepackage{courier}  % DO NOT CHANGE THIS
\usepackage[hyphens]{url}  % DO NOT CHANGE THIS
\usepackage{graphicx} % DO NOT CHANGE THIS
\usepackage{natbib}  % DO NOT CHANGE THIS AND DO NOT ADD ANY OPTIONS TO IT
\usepackage{caption} % DO NOT CHANGE THIS AND DO NOT ADD ANY OPTIONS TO IT
\usepackage{amsmath}
\usepackage{amssymb}
\usepackage{mathrsfs}
\usepackage{tikz}
\usepackage{tikz-qtree}
\usepackage{multirow}
\usepackage{soul}
\usepackage{listings}
\usepackage{algorithm}
\usepackage{algpseudocode}
\usepackage{algorithmicx}
\usepackage[english]{babel}

\usepackage[linewidth=0.5pt]{mdframed}
\usetikzlibrary{positioning}
\usetikzlibrary{trees}
\usetikzlibrary{decorations.pathreplacing,calc}
\usetikzlibrary{shapes.misc}
\usetikzlibrary{calc,shapes.multipart,chains,arrows}

\newcommand{\mycomment}[1]{}

\newlength\myindent
\title{Generating Concurrent Programs From Sequential Data Structure Knowledge}
\author {
        Sarat Chandra Varanasi,\textsuperscript{\rm 1}
        Neeraj Mittal, \textsuperscript{\rm 1}
        Gopal Gupta \textsuperscript{\rm 1} \\
}
\affiliations {
    \textsuperscript{\rm 1} Department of Computer Science, \\
    The University of Texas at Dallas, \\
     Richardson, Texas, USA \\
    sarat-chandra.varanasi@utdallas.edu, 
    neerajm@utdallas.edu, gupta@utdallas.edu
}
\begin{document}

%\begin{abstract}
%    We present a novel technique using human style informal reasoning to transform knowledge from Sequential Data Structure Operations into their equivalent Concurrent versions. Transforming sequential programs to safe concurrent code is a non-trivial task even for programs as small as two lines of code. Our technique highlights the effectiveness of informal reasoning by simulating the proof steps performed by a human expert using a knowledge base of a given sequential data structure. Human style proof steps involve deduction, induction and abduction using a background theory and knowledge. Answer Set Programming (ASP) allows for succint modelling of First order theories of Pointer Data Structures. Combined with the background theory of given pointer data structure and knowledge about its sequential operations, our reasoner can systematically make the same judgments a human reasoner while constructing a proof of safe concurrent code. Several reasoning challenges involved in order to transform the sequential data structure into its equivalent concurrent version are presented. All these reasoning tasks are encoded in ASP and our reasoner can make sound judgments towards concurrent data structure transformation. To our knowledge, our work is a unique attempt and the first step towards transforming sequential programs into concurrent versions using a background theory and knowledge base. 
%\end{abstract}

\maketitle

\begin{abstract}
In this paper we tackle the problem of automatically designing concurrent data structure operations given a sequential data structure specification and knowledge about concurrent behavior.  Designing concurrent code is a non-trivial task even in simplest of cases.  Humans often design concurrent data structure operations by transforming sequential versions into their respective concurrent versions.  This requires an understanding of the data structure, its sequential behavior, thread interactions during concurrent execution and shared memory synchronization primitives. We mechanize this design process using automated commonsense reasoning. 
We assume that the data structure description is provided as axioms alongside the sequential code of its algebraic operations. This information is used to automatically derive concurrent code for that data structure, such as dictionary operations for linked lists and binary search trees. Knowledge in our case is expressed using Answer Set Programming (ASP), and we employ deduction, induction and abduction---just as humans do---in the reasoning involved.
ASP allows for succinct modeling of first order theories of pointer data structures, run-time thread interactions and shared memory synchronization.  Our reasoner can systematically make the same judgments as a human reasoner while constructing provably safe concurrent code. We present several reasoning challenges involved in transforming the sequential data structure into its equivalent concurrent version. All the reasoning tasks are encoded in ASP and our reasoner can make sound judgments to transform sequential code into concurrent code. To the best of our knowledge, our work is the first one to use commonsense reasoning to automatically transform sequential programs into concurrent code.
\end{abstract}

\section{Introduction}
    We present a novel technique that generates concurrent programs for pointer data structures given a first order (logic) data structure theory and background knowledge about its sequential operations. Design of concurrent operations for data structures is non-trivial. As we show in this paper, there are several challenges that need to be addressed given a data structure description. Traditionally, concurrent programs are designed manually and their proofs of correctness are done by hand. Few concurrent data structures are also verified using symbolic bounded model checking \cite{vechev2010deriving}. Avoiding state space explosion in the verification of concurrent programs is the main challenge for symbolic model checkers. Several works address this issue in interesting ways \cite{emerson2000reducing, vechev2008deriving}. Other formal approaches involve performing Hoare-Style Rely-Guarantee reasoning \cite{vafeiadis2006proving} to verify concurrent programs that have been manually designed. 
    %GG:
    These approaches, thus, seek help of automated verification in an otherwise manual design process to ensure correctness. In contrast, our approach leverages reasoning techniques employed in AI,  knowledge about concurrency, and explicitly modeled sequential data structure code to arrive at a safe concurrent program. Work in model checking and formal logics for concurrency do not exploit the sequential data structure knowledge. Their main focus is to prove absence of incorrect thread interactions (or traces). The proof of correctness of the verified concurrent code is provided outside their frameworks, assuming certain symmetry properties on concurrent interactions. Our work, in contrast, performs the reasoning tasks that an expert in concurrent program design deliberates in order to construct a safe concurrent program. This requires an understanding of the data structure representation, the library of algebraic operations that modify the data structure, an understanding of shared memory and how primitive read and write operations affect the shared memory. Additionally, the expert can explicitly describe the safety conditions that are desired, the invariants that need to be preserved during concurrent execution. With this knowledge, the expert obtains the concurrent program that acquires the ``right" number of locks (synchronization steps) that is safe for any concurrent interaction with an unbounded number of threads.   
    
    Our work can be seen as applying automated (commonsense) reasoning \cite{mccarthy1960programs} to the program synthesis problem. To the best of our knowledge our is the first effort that attempts to emulate the mind of a human domain expert who designs concurrent data structure using sequential ones as a starting point. 
\section{Background}
  %\par
  \subsection{Answer Set Programming}
    Answer Set Programming is a declarative problem solving paradigm with applications spanning several areas of AI research: from planning to complex human style commonsense reasoning \cite{erdem2016applications,chen2016physician}. The expressive power of ASP is due to its non-monotonic reasoning capabilities. Non-monotonic reasoning allows one to retract conclusions in light of new evidence. 
    %Gopal2020: explain what nonmon is?
    Fundamentally, ASP programs are normal logic programs with non-monotonicity, i.e., we can write code in ASP to take an action if a proof fails. This is achieved through support for negation as failure. This helps model commonsense reasoning (humans can take an action predicated on failure of a proof). Monotonic logics cannot reason about proof-failure within the logic itself. 
    
    An ASP Program consists of rules of the form $\{p \leftarrow q_1, q_2,..q_i, not \ r_1, not \ r_2, .., \ not \ r_j\}$. If $i = j = 0$, then $p$ is a fact. If $p$ is the empty ($\square$ or false), the rule represents a constraint. The operator $not$ represents negation-as-failure. The set of satisfiable literals of an answer set program are termed as its answer sets (stable models). A stable model $m$ is entailed by an ASP program $\Pi$, ie., $\Pi \models m$ if and only if $m$ is present in every stable model of $\Pi$ \cite{faber2009manifold}. We assume basic familiarity with stable model semantics and ASP solvers \cite{gebser2016theory}. More details about ASP can be found elsewhere \cite{gelfond1988stable,gelfond2014knowledge}.
   % \par
  \subsection{Concurrent Data Structures}
    Concurrent Data Structures usually support data structure dictionary operations being manipulated by an unbounded number of interacting threads. They are nothing but multiprocessor programs.  We only assume a sequentially consistent shared memory model in this paper. Sequential consistent memory allows any update performed on the shared memory to be visible, before performing a subsequent read, to every thread in the system. Concurrency can be viewed as a sequence of interleaved steps taken by various threads in the system. A concurrent program is the set of interleaved traces it generates. 
     To  make sense of correctness of concurrent data structures, the notion of \textit{linearizability} \cite{herlihy2011art}is widely used. A concurrent data structure is termed linearizable, if the effects of concurrent modification by several threads can be viewed as if the concurrent operations were performed in some sequential order. In this paper, we study the modifications performed on a data structure as if they are respecting a serialized schedule. This allows us to model concurrency in an intuitive manner and sidesteps the necessity to understand traces. This assumption is sufficient to generate safe concurrent programs. However, to guarantee deadlock-free programs, one needs to also analyze traces. %do we need to worry about deadlock-freedom? omit the sentence? 
\section{General Notions}
   \subsection{Data Structures}
   
    %\subsection{Few Data Structure Notions}
    \noindent     Data structures include some representation of information and the dictionary operations associated with them such as membership, insert and delete. Representation itself involves several notions at various levels of abstraction. For example, to describe a linked list, one needs primitive notions of \textit{nodes} contained in memory, connected by a chain of \textit{edges}. Further, there are notions of \textit{reachability} (or \textit{unreachability}) of nodes and keys being present (or \textit{absent}) in a list. Membership operations usually involves traversing the elements (or nodes) in the data structure until an element(s) satisfying certain criteria is found. Insert operation also involves traversing the data structure until a right ``window" of insertion is found. Similarly, the delete operation removes the appropriate elements in a certain window. The notion of window represents some local fragment of the data structure that is modified as part of a data structure update operation. This notion is useful when discussing about locking nodes in concurrent programs.
   % \par  
    \subsection{Tree-Based Pointer Data structures}
     A heap is a collection of nodes connected by edges. A data structure $\mathcal{D}$ is a recursive definition defining a tree of nodes in memory. Further, the only primitive destructive operation that may be performed is linkage of pointers: $link(x,y)$. The abstract relation $link(x,y)$ links node $y$ to $x$ in the heap. We support transformation of concurrent code for an algebraic operation $\sigma_{\mathcal{D}}$ associated with $\mathcal{D}$ such that $\sigma_{\mathcal{D}}$ may be performed in a constant number of $link$ operations. For instance insert operation for linked lists can be performed in two steps.
    % \par
    \subsection{Data Structure Theory and Knowledge}
  We assume that a first order theory $\mathcal{T}$ is provided for a pointer data structure along with the sequential data structure knowledge $\mathcal{K}$. We use the theory for linked lists and its knowledge as running example in this paper. The technique however applies to all tree-based pointer data structures. The theory and knowledge are provided in Figures 1 and 2. The data structure theory defines linked lists as a chain of edges with special sentinel nodes $h$ at the head of the list and $t$ at the end. The meaning of predicates $reach$ and $present$ is straightforward.
  
  %\par 
  \subsection{Sequential Data Structure Knowledge}
The knowledge $\mathcal{K}$ contains the pre/post-conditions of insert and delete operations for linked lists. The primitive read and write steps are captured by \textit{deref} (dereferencing pointer) and \textit{link} (link-pointer) operations. The effects of link operation are also described using \textit{causes} relation. The knowledge $\mathcal{K}$ is useful for two purposes: 1. It bounds the interference effects of arbitrary thread interactions in a concurrent execution. 2. It narrows the blocks that need to be synchronized to obtain a concurrent algorithm. However, as we present next, there are several challenges to transforms Steps $\langle1,2\rangle$ of \textit{insert} operation into a concurrent version.  
  The program statements are encoded within the vocabulary of the data structure using answer set programming (ASP). Program Blocks in computer programs can be viewed as equivalence class of input-output transformation.
   Further, the program blocks perform destructive update operations on the data-structure (insert/delete). 
   Given the data structure definition, it is straightforward to generate data structure instances that satisfy a given equivalence class. This is because the assumed data structure definition $\mathcal{D}$ is recursive. The recursive definition can enumerate the set $S_{\mathcal{D}}$ of all structurally isomorphic instances of $\mathcal{D}$. The set $S_{\mathcal{D}}$ can be ordered by the number of recursion unfoldings used to generate the instances, starting from the least number of unfoldings. For $\mathcal{D}_i, \mathcal{D}_j \in S_{\mathcal{D}}$, $i < j$ implies that $\mathcal{D}_i$ is a ``smaller" structure than $\mathcal{D}_j$ and appears before $\mathcal{D}_j$ in the recursion depth ordering.  
 
     \setlength{\fboxsep}{1pt}
   \begin{figure}
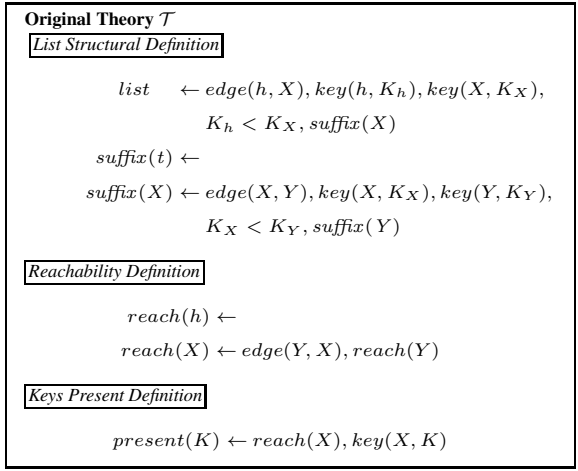
 
   \fbox{\scriptsize
                            \begin{tabular}{p{7cm}}
                            {\bf 
                             Original Theory $\mathcal{T}$}  \\
                           {                
                            \begin{minipage}{0.97\linewidth} 
                             \textit{ \fbox{\textit{List Structur}al Definition}} 
                             \begin{align*}
                             list \phantom{fo} \leftarrow & ~edge(h, X), key(h, K_h), key(X, K_X), \\
                             \phantom{thequickbrownfox}  &  ~K_h <  K_X, \mathit{suffix(X)}
                             \\
                             \mathit{suffix(t)} \leftarrow & \\
                             \mathit{suffix(X)} \leftarrow & ~edge(X, Y), key(X, K_X), key(Y, K_Y), \\
                             \phantom{thequickbrownfox}    & ~K_X < K_Y, \mathit{suffix(Y)} 
                             \end{align*} 
                             \fbox{\textit{Reachability Definition}} 
                          \begin{align*} 
                            reach(h) \leftarrow & \\
                            reach(X) \leftarrow & ~edge(Y, X), reach(Y) 
                            \end{align*} 
                            \fbox{\textit{Keys Present Definition}} 
                            \begin{align*}
                            present(K) \leftarrow & ~reach(X), key(X, K)
                            \end{align*}
                            \end{minipage}
                           }
                            \end{tabular}
                            }                  
     \label{fig:Theory}
     \caption{Linked List Theory}
     \end{figure}

\section{Challenges in Transforming Sequential Data Structures To Concurrent Code}
     We assume that the traversal code remains the same as the sequential version for a lock-based concurrent data structure. 
     %Gopal2020: what does previous sentence mean?
     Therefore, the challenges we discuss are purely for destructive update program steps. We present the challenges involved and how they are addressed in turn. \par
    
    %Gopal2020: do we need to explain that there is special sentinel node and an end node? Also, sequential linked list knowledge may have to be explained in a few sentences.
  
    \subsection{Order of the Program Steps Matter}
        Consider the task of inserting a single node in a linked list. The linked list insert operation can be carried out using the following two statements executed in order: 1. \texttt{x.next := target} 2. \texttt{target.next := y}. However, this order of pointer linkages is undesirable in a concurrent setting, particularly when considering a concurrent membership test occurring on the list.  This is because after executing step 1, the list is broken. That is, the chain of connected nodes in the list ends with node \textit{target}. Therefore, node \textit{y} is unreachable with respect to a concurrently executing membership test. Thus, to make a concurrent membership test to only observe a well-formed list at all times, one needs to preserve invariants specified for a concurrent execution. These invariants can be easily encoded in logic. For example, we can write the following constraint in ASP: $\{\leftarrow reach(X, 0), not \ reach(X, T)\}$. This constraint says a node can never become unreachable when executing the insert operation. Thus, the order of steps $\langle1, 2\rangle$ is rejected. The correct order for concurrent setting is $\langle2, 1\rangle$. \par
        In general, it is possible that there exists no ordering of steps that preserves the invariant in a concurrent execution. Then, the designer uses the Read-Copy-Update (RCU) \cite{mckenney2013rcu} technique to copy the window and perform changes locally (outside shared memory) and atomically splice window back to the shared memory. The RCU technique depends on the ability to splice back the window atomically. For tree-data structures, if the window is a sub-tree, then it is easy to atomically splice a sub-tree to shared memory by updating its parent pointer in the shared memory. The applicability of RCU framework can be either made explicit in the data structure knowledge, or should otherwise be inferrable from the knowledge of data structure representation/operations.   
   % \par
        \begin{figure}
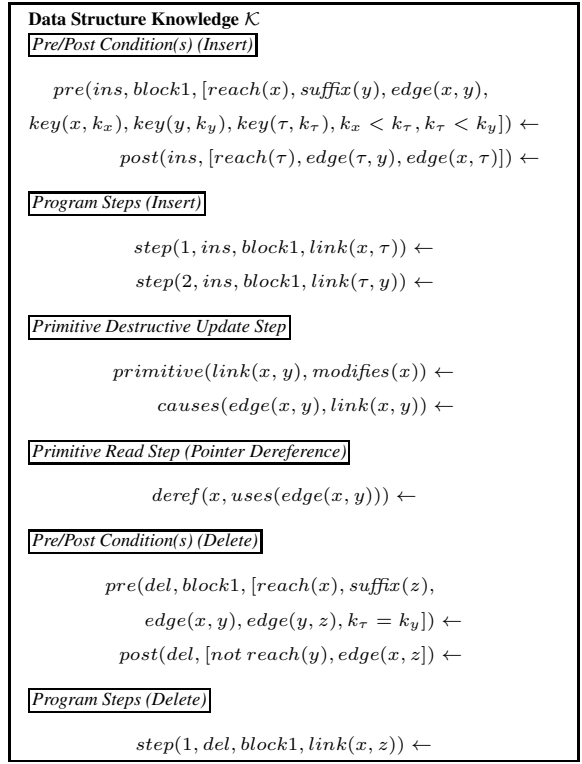

   \fbox{\scriptsize
                            \begin{tabular}{p{7cm}}
    \bf{Data Structure Knowledge} $\mathcal{K}$ \\
                             {
                             \begin{minipage}{0.97\linewidth}
                             \fbox{\textit{Pre/Post Condition(s) (Insert)}} 
                             \begin{align*}
                              pre(ins, block1, [reach(x), \mathit{suffix(y)}, edge(x, y), \phantom{th} & \phantom{\leftarrow} \\ 
                             key(x, k_x), key(y, k_y), key(\tau, k_\tau), k_x < k_{\tau}, k_{\tau} <  k_y])   & \leftarrow  \\
                             post(ins, [reach(\tau), edge(\tau, y), edge(x, \tau)]) & \leftarrow
                             \end{align*}
                             \fbox{\textit{Program Steps (Insert)}} 
                              \begin{align*}
                               step(1, ins, block1, link(x,\tau)) & \leftarrow\\
                               step(2, ins, block1, link(\tau, y)) & \leftarrow
                              \end{align*}        
                             \fbox{\textit{Primitive Destructive Update Step}} 
                              \begin{align*}
                                   primitive(link(x,y),\mathit{modifies(x)}) & \leftarrow \\
                                   causes(edge(x,y),link(x,y)) & \leftarrow 
                               \end{align*}
                              \fbox{\textit{Primitive Read Step (Pointer Dereference)}} 
                              \begin{align*}
                                \mathit{deref}(x, uses(edge(x,y))) & \leftarrow 
                             \end{align*}
                             \fbox{\textit{Pre/Post Condition(s) (Delete)}} 
                             \begin{align*}
                             pre(del, block1, [reach(x), \mathit{suffix(z)},  & \phantom{\leftarrow}
                             \\ 
                             edge(x, y), edge(y, z), k_\tau = k_y]) & \leftarrow  \\
                                post(del, [not \ reach(y), edge(x, z]) & \leftarrow
                             \end{align*}
                             \fbox{\textit{Program Steps (Delete)}} 
                              \begin{align*}
                               step(1, del, block1, link(x,z)) & \leftarrow
                               \end{align*}
                            
                             \end{minipage}
                             }
                             \end{tabular}
                             }
    \label{fig:Knowledge}
    \caption{Sequential Linked List Knowledge}
    \end{figure}
  
  \medskip  
    \noindent\textbf{Lock Acquisition:}
           When transforming sequential code to concurrent code, a designer has to determine which memory cells to lock in order to ensure a safe execution. An initial strategy is to acquire all the locks involved in the destructive update steps. This is reasonable as the steps explicitly describe the nodes that are modified in the computation. However, in practice, acquiring the right number of locks is hard. This is true for Internal BSTs where the window of modification is not contiguous and depends on the size of the sub-tree that is being modified. Therefore, the correct lock acquisition strategy must be inferred given the data structure axioms. Also, acquiring the minimal number of locks is desired as this reduces the synchronization overhead and improves the overall throughput of the concurrent execution. 

    %\par
\subsection{Concurrent Traversal may require RCU}
As we mentioned before, our assumption is that transformation for concurrent membership operations is vacuous (i.e., membership operation is unchanged in a concurrent setting). This ensures that the membership queries execute as fast as possible while acquiring no locks. However, for the membership operation to work consistently, the code for insert and delete operations should work correctly. We illustrate this with an Internal BST example. 
          Consider the following internal BST shown and assume a thread is about to delete the node
          \textit{l}. The inorder successor of \textit{l} is \textit{lrl}. It is clear that the delete operation should lock all the nodes on the path from \textit{l} to \textit{lrl} (inclusive). However, this locking scheme is inadequate although it modifies the data structure in a consistent manner. The problem lies outside the code of delete operation itself. The problem surfaces with a concurrent membership operation looking for node \textit{lrl}. Due to node (and henceforth key) movement, it is possible for the traversal code to miss $lrl$. Again, this scenario needs to be inferred from the data structure knowledge. Due to arbitrary key-movement, a concurrent membership operation might claim that it does not see a node as part of the data structure when it is conceptually part of the structure. This scenario can also be addressed by the RCU framework.  
   %  \par
\subsection{Proving  correctness of a Concurrent Algorithm} 

Proving that a set of algebraic operations are thread-safe may involve several proof obligations in general. For linearizable data structures, it is sufficient to show that every execution of the insert, delete and membership operations is equivalent to some serialized execution. This implies that the pre/post-condition invariants associated with the sequential algorithm are never violated in any concurrent execution. That is, when a thread is modifying the data structure with respect to an algebraic operation, it is the only agent in the system modifying that fragment of the data structure necessary to complete the algebraic operation. A domain expert who does these proofs by hand in practice, identifies all the destructive update steps performed by each operation. Then, he/she ensures that if the correct shared memory variables are locked, then any potential interference from other operations does not violate the invariants associated with the destructive update steps.  
   Therefore, our reasoner would perform these proof obligations in a way a domain expert would, given the data structure knowledge and representation.

\section{Transforming Sequential Data Structures to Concurrent Data Structures}

\mycomment{
\subsection{Pre-condition Invariants}
   Consider the task of generating the concurrent code for Linked List insert, delete and membership (vacuous) operations given the respective sequential code and background theory of linked lists. Both insert and delete operations involve traversing through the nodes of the data structure until the right precondition for destructive update is met. Again, we assume that no change is required for traversal part of insert (delete) operation. We are only concerned with acquiring the correct locks and executing the destructive update steps in the right order to preserve the necessary invariants. The preconditions for insert and delete operations are respectively $pre_{ins}$ and $pre_{del}$ from the data structure knowledge $\mathcal{K}$.}

 \mycomment{  The conditions are given below:
  \fbox{
     \begin{minipage}{0.97\linewidth}
      $pre_{ins}(x, y, \tau) \equiv \{reach(x), edge(x, y), \mathit{suffix(y)}, \\
       key(\tau, k_{\tau}), key(x, k_x), key(y, k_y), k_x < k_{\tau}, k_{\tau} < k_yan\}$ \\
       $pre_{del}(x, y, \tau) \equiv \{reach(x), edge(x, y), edge(y, z), \\ \mathit{suffix(z)}
       key(\tau, k_{\tau}), key(x, k_x), key(y, k_y), eq\_key(k_y, k_{\tau})$
     \end{minipage}  
  }}
 %  \par
   \subsection{Modeling Thread Interference}
      Interference is simply arbitrary mutations that might occur on the data structure when some thread is observing the data structure. A domain expert when proving the correctness of insert operation, would consider the code for the insert operation as an agent trying to insert a target key while being aware of arbitrary changes that the environment might perform. The changes that the environment might perform can be quite arbitrary. However, given that the only \textit{effectful} operations in the concurrent execution are insert and delete, the domain expert assumes the instantaneous effects of an insert or delete operation when an agent is performing its own steps. To make this concrete, the only possible destructive update effects by the environment are either the effects of insert or delete operation. Given the sequential data structure knowledge of the insert and delete operation precondition and their effects, the interference model instruments an environment agent that picks arbitrary nodes from the data structure or nodes from the heap extraneous to the data structure and performs instantaneous edits (algebraic mutations). We argue that this model is sufficient to discover any undesired thread interactions. The sufficiency of the interference model stems from reasoning interference effects based on sequential algorithm equivalence classes. This feature is usually not present in a concurrent program verification task performed via model checking. However, model checkers may also be instrumented with additional abstractions to guide their search for counterexample traces\cite{vechev2010abstraction}  
       Also, a domain expert would informally follow this line of argument when arguing for correctness. 
  %     \par
                       \setlength{\fboxsep}{1pt}

  \begin{center}
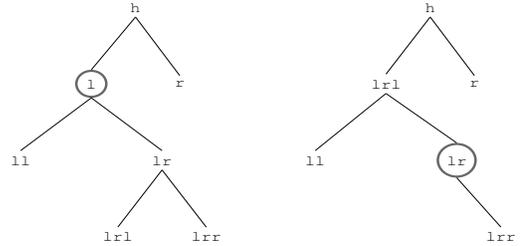

 \resizebox{7cm}{3.3cm}{
\begin{tikzpicture}[
roundnode/.style={circle, draw=black!60,  very thick, minimum size=3mm},
squarednode/.style={rectangle, draw=black!60,  very thick, minimum size=5mm},
-
]
   \fontsize{8}{9}\selectfont\ttfamily
   \tikzstyle{level 1}=[distance=1mm,sibling distance=15mm]
  \tikzstyle{level 2}=[distance=1mm,sibling distance=24mm]
  \tikzstyle{level 3}=[distance=1mm,sibling distance=15mm]
  
  \draw (-5,0) node  {h}
    child {node[roundnode] {l}
             child {node {ll}}
             child {node {lr}
              child {node {lrl}}
              child {node {lrr}}}
             }
    child {node {r}};

    \draw (0,0) node  {h}
    child {node {lrl}
             child {node {ll}}
             child {node[roundnode] {lr}
               child [missing]
               child {node {lrr}}
             }
              }
    child {node {r}};
     
\end{tikzpicture} 
 } 
 \captionof{figure}{Traversal operation reaches till node $l$ but misses $lrl$ by the time it dereferences $l.right$}
\end{center}
  
    \subsection{Predicate Falsification in Concurrent Execution}
       To preserve invariants of a sequential execution in a concurrent setting, it is necessary to know the predicates that can be falsified in a concurrent execution. Since the sequential data structure knowledge provides the necessary preconditions, we systematically check for potential falsification of every conjunct in $pre_{ins}$ (or $pre_{del}$) with respect to environment interference. Predicate $pre\_{ins}$ (and similarly $pre_{del}$) is defined as $\{pre_{ins}(X,\tau,Y) \leftarrow reach(X), \mathit{suffix(Y)}, edge(X,Y), k_X < k_{\tau} < k_Y\}$ which is picked from the third argument of $pre(ins,block1,..) \in \mathcal{K}$ . If a predicate is not falsified with respect to the interference model, then it is indeed not falsifiable in \textit{any} serialized concurrent execution. This implies, that a thread need not synchronize on the un-falsifiable predicate(s). For example, in $pre_{ins}$ the predicate $\mathit{suffix(y)}$ is not falsifiable. This is because, any correct algebraic mutation would only skip the node $y$ but not unlink it in the chain to tail node $t$. 
  % \par
   \subsection{Lock Acquisition from Critical Conditions} 
         Locks are necessary to protect the invariant predicates from falsification by interference. A conservative approach is to associate locks with every predicate and acquire locks. This approach can be taken for general concurrent programs where less semantic knowledge is available about the sequential program that is being transformed \cite{deshmukh2010logical}. In a fine-grained locking scheme, the only locks that can be acquired are the set of reachable nodes of the pointer data structure.  Intuitively, locking the nodes involved in the window of modification seems sufficient. Although locking this set of nodes is insufficient in general, this sets a lower bound on the number of locks to acquire in a fine-grained locking scheme. Once, the right set of nodes to be locked is guessed, the domain expert confirms the non-falsifiability of the invariants. If non-falsifiability is affirmative, then the concurrent program would only acquire the guessed locks. 
   % \par
   
  \mycomment{ \subsection{Lazy Synchronization}
      Guessing the right locks alone is not sufficient to generate a correct algorithm. Because acquisition of multiple locks is not an atomic program step, there might be invariants that may be falsified if other threads are operating on the data structure. Hence, the concurrent program validates the pre-condition after acquiring all the locks. This validation can be performed non-atomically, as no other thread can falsify the invariant when locks have been acquired. If the validation fails, then the thread releases the locks and aborts. This technique is known as Lazy Synchronization \cite{herlihy2011art} as the thread optimistically acquires locks and post-checks the invariants later.
      }
 
  \section{Decomposition of Concurrency Proof Obligations into Reasoning Tasks}
  % \par
   \subsection{Generating Interference Model}   
       We assume a theory $\mathcal{T}_{\mathcal{D}}$ encodes the structural definition of $\mathcal{D}$ along with various primitives and abstractions necessary to understand conditions and effects involved in arbitrary manipulations that might be performed on instances of $\mathcal{D}$. It is also assumed that we can identify predicates that are time invariant from the predicates that are time dependent. Theory $\mathcal{T}^{\mathcal{R}}$ is the planning domain \cite{lifschitz2019answer} with reified time argument. The theory $\mathcal{T}^{\mathcal{R}}$ contains all the predicates that are time dependent with an extra argument for time. More precisely, for all $p(\bar{X}) \in \mathcal{T}$ that is time dependent, $p(\bar{X}, T) \in \mathcal{T}^{\mathcal{R}}$.
       Also, the ordering of time is captured by the $next \in \mathcal{T}^{\mathcal{R}}$ relation, where $next(t, t')$ implies time step $t'$ follows after $t$. The $next$ relation is transitive. 
       \mycomment{An example for linked list insert operation is provided:
       \fbox{
       \begin{minipage}{0.97\linewidth}
         $step(1, ins, link(x,\tau)). \ step(2, ins, link(\tau, y)).$
         $\\ pre(ins,[reach(x),edge(x,y),\mathit{suffix(y)}, 
       key(x,k_x), \\ key(y,k_y),key(\tau,k_{\tau}), k_x < k_{\tau}, k_{\tau} < k_y]).$. 
        $\\ post(ins,[reach(\tau),edge(x,\tau),edge(\tau,y)]).$. 
        \end{minipage}
        }}
        From the procedural information in $\mathcal{K}$, it is easy to model the instantaneous effects of the actions. Let the theory encoding the interference model be represented as $\mathcal{I}$. For operation $ins$, a predicate $\mathit{interfere(ins,x,\tau,y)}$ is added to $\mathcal{I}$ as an abducible. Abducibles are predicates that are guessed or falsified non-monotonically in ASP.  Note that $pre\_{ins}$ contains exactly the same terms in the data structure procedural knowledge replaced with uppercase variables. Also, \textit{interfere} is modeled as an abducible, to give the interference model enough flexibility for other interference operations to take place. \\
        \par\noindent
        {
         \fbox{
         \begin{minipage}{0.97\linewidth}
         $abduce  \ \mathit{interfere(ins,X,\tau,Y,T)} \\ $ 
         $abduce \ \mathit{interfere(del,X,\tau,Y,T} \\$
         $\mathit{interfere(ins,X,\tau,Y,T)} \leftarrow  pre\_ins(X,Y,\tau,T) \\$ 
         $\mathit{interfere}(del,X,\tau,Y,T) \leftarrow pre\_del(X,Y,\tau,T)$
         \end{minipage}
         }
         }
         \par\noindent
         Now, the two effects are also encoded as properties following from \textit{interfere}. Because Normal Logic Programs are (almost) Horn Clauses, we allow at most one positive literal in the head.
         The two effects are therefore encoded as two different consequences of \textit{interfere}.
         \par\noindent \\
         \fbox{
         \begin{minipage}{0.97\linewidth}
         $edge(X, \tau,T') \leftarrow \mathit{interfere}(ins, X, \tau, Y,T), next(T, T')\\$
         $edge(\tau, Y,T') \leftarrow \mathit{interfere}(ins,X,\tau, Y, T), next(T, T') \\ $
         $edge(X, Y, T) \leftarrow \mathit{interfere}(del,X,\tau,Y,T), next(T, T')$ 
         \end{minipage}
         }
         \par\noindent \\
        Similarly for every algebraic operation $\sigma \in \mathcal{K}$ an interference predicate $\mathit{interfere_{\sigma}}$ is added to $\mathcal{I}$. 
       % \par
        \subsection{Checking Falsification Predicates (Task 1)}
        Given the data structure knowledge $\mathcal{K}$, one can discharge conditions that check for falsifications of every conjunct in $pre_{\sigma}$. For $pre_{ins}$ of insert operation in linked list, one can generate predicate falsification checks for \textit{reach, suffix,} and \textit{edge}. These are precisely the time dependent relations in $\mathcal{T}$. The predicate \textit{falsify\_reach} checks for falsification of reach in one time step. The definition looks like: $\mathit{\{falsify\_reach}  \leftarrow reach(X, T), not \ reach(X, T'), next(T, T')\}$. Similarly the falsification of \textit{suffix} and \textit{edge} are defined. These falsification predicates are also added to $\mathcal{I}$. 
        Once the falsification predicates are added to the theory $\mathcal{I}$, one can check if the falsification predicates are true in some model of $\mathcal{T}^{\mathcal{R}} \cup \mathcal{I}$. If their satisfiability is affirmative, then interference indeed falsifies the predicates. Otherwise, the interference cannot falsify the predicates. This check for falsification is an optimization step in order to reduce the number of predicates to be validated post lock-acquisition.
      %   \par
        %\subsection{Guessing Locks}
          
      %  \par
        \subsection{Checking Adequacy of Guessed Locks (Task 2)}
          From the procedural knowledge of the data structure operations, it is easy to guess the locks to be acquired. As an initial guess, every thread should at least synchronize on the nodes involved in the ``window" of modification. For example, the window for insert w.r.t $pre\_ins$ is the set of nodes $\{x,y\}$.
           \mycomment{Once the set of ``window" nodes have been identified, then the concurrent algorithm would only acquire locks on these nodes. For instance, in linked list insert concurrent code, the only locks that would be acquire are on nodes $\{x,y\}$. Another approach is to recursively acquire all the locks that are reachable from nodes $\{x, y\}$.}
            After guessing the set of locks to be acquired, one can now check their adequacy in the presence of interference. In the interference model, the effects of \textit{interfere} predicates are enabled only if there are no locks already acquired on the nodes they modify. For instance, for $\mathit{interfere}(ins,X,\tau,Y)$  the nodes that are modified are $\{X, \tau\}$ (for arbitrary $X$). Both the two effects shown previously, are enabled only when there are no locks on $X$ or $\tau$. These re-written rules are part of the theory $\mathcal{I}^{\mathcal{L}}$ which represent the reified interference model in the presence of locks. The re-written rules are shown below: \\
            
             \fbox{
             
         \begin{minipage}{0.97\linewidth}
         $edge(X, \tau,T') \leftarrow \mathit{interfere}(ins, X, \tau, Y,T), next(T, T'), \\
                        \hphantom{thequickbrownfo} not \ locked(X, T)$
         $\\edge(\tau, Y,T') \leftarrow \mathit{interfere}(ins,X,\tau, Y, T), next(T, T'),\\ 
                        \hphantom{thequickbrownfo} not \ locked(\tau,T)$ 
         $\\edge(X,Y,T') \leftarrow \mathit{interfere}(del, X, \tau, Y, T), next(T, T') \\
                        \hphantom{thequickbrownfo} not \ locked(X, T)$
         \end{minipage}
         }
         \\ \\ 
         %\par\noindent 
           The locked nodes themselves are captured by the \textit{locked} relation and are added as facts to $\mathcal{I}^{\mathcal{L}}$. The falsification predicates remain the same in $\mathcal{I}^{\mathcal{L}}$. From $\mathcal{I}^{\mathcal{L}}$ , one can infer entailment of the falsification predicates. If the answer is affirmative, then the locking scheme is clearly inadequate. Otherwise, the locking scheme is adequate and the concurrent code can be generated (with lazy synchronization). If the locking scheme is inadequate, the reasoner might use another locking scheme or can recommend the RCU framework for synchronization.
           \mycomment{If the locking scheme is adequate, a final decision to be made is to check if the operation results in node (key) movement within the data structure. If the answer is yes, then again the reasoner recommends RCU. Otherwise, the generated concurrent code is correct.}
      %    \par
          \subsection{Validating Sequential Program Order (Task 3)}
              We denote $pre(\sigma)$ and $post(\sigma)$ to be the pre-condition and post-condition associated with operation $\sigma$.
              Given the sequential program steps in $\mathcal{K}$, one should be able to infer the right order of program steps that do not violate an invariant in a concurrent setting. A common invariant that needs to be satisfied is the well-formedness of the data structure at all times. Having the data structure well-formed at all times is desirable as it makes the results returned by membership queries easier to explain with respect to linearizability. Given an invariant $Inv$, theory $\mathcal{T}^{\mathcal{R}}$ and procedural knowledge in an operation $\sigma$ in knowledge base $\mathcal{K}$, a new theory $\mathcal{T}^{po}$ can be generated that validates the program order of all basic blocks with respect to invariant $Inv$. For every program step $s_{\sigma}(\bar{X}) \in K$, a reified abducible is $s_{\sigma}(\bar{X}, T)$ is generated and added to $\mathcal{T}^{po}$. Then, the $post(\sigma)$ is reified and added to $\mathcal{T}^{po}$. Similarly $Inv$ is also added to $\mathcal{T}^{po}$. Also, the necessary time steps along with their ordering using $next$ is added to $\mathcal{T}^{po}$. Now, if $\mathcal{T}^{po}$ is satisfiable, then there exists a program order that does not violate the $Inv$ throughout. The order might be permuted. Because, the program steps are bounded by a constant number $n$, the original program steps can be given unique names to map the original program order to the permuted order.   
              On the contrary, if $\mathcal{T}^{po}$ is unsatisfiable, then no permutation (including the original program order) exists that can preserve $Inv$. In that case, the reasoner would recommend using an RCU Synchronization. 
               \begin{figure}
              {
              \fbox{
                % \small
                \begin{minipage}{0.97\linewidth}
                 \textit{Original Traversal Code:} \\
                 $traverse(X) \leftarrow edge(X, Y), not \ pre\_ins(X, \tau, Y), \\ traverse(Y)$ \\
                 $traverse(X) \leftarrow edge(X, Y), pre\_ins(X, \tau, Y)$ \\
                 \textit{Instrumented Async Observer:} \\
                 $traverse(X, T) \leftarrow edge(X, Y, T), \\ not \ pre\_ins(X, \tau, Y, T), traverse(Y, T'), next(T, T')$ \\
                 $traverse(X, T) \leftarrow edge(X, Y, T), pre\_ins(X, \tau, Y, T)$  \\
                 \textit{Oracle Observer:} \\
                 $traverse^{\mathcal{O}}(X, T) \leftarrow edge(X, Y, T), \\ not \ pre\_ins(X, \tau, Y, T), traverse^{\mathcal{O}}(Y, T)$ \\
                 $traverse^{\mathcal{O}}(X, T) \leftarrow edge(X, Y, T), pre\_ins(X, \tau, Y, T)$ \\
                 \textit{Key Movement Predicate:} \\
                 $keymove \leftarrow traverse^{\mathcal{O}}(X, T), not \ traverse(X, T)$
                \end{minipage}
              } 
              }
              \label{fig:traversal}
              \caption{Modeling Missed-Key Scenario}
              \end{figure}
       %   \par
          \subsection{Detecting Key-Movement (Task 4)}
              This task is necessary to tackle the keys missed by a concurrent traversal operation as shown for Internal BSTs. 
              We detect key-movement based on the differences in the set of observed keys by observed an asynchronous observer (traversal code) and a synchronous observer (oracle) in the presence of interference. The traversal code from $\mathcal{K}$ can be instrumented to determine the exact set of nodes (keys) visited by the traversal in the reified interference model $\mathcal{I}$. We assume the traversal code is recursive and the pointer dereferences are identifiable using $\mathcal{K}$. The instrumented predicate simulates the passage of one time step after every dereference. This enables the interference model to make simultaneous changes and alter what might be otherwise observed by the traversal code (in the absence of interference). 
              At the same time, the oracle observer is a predicate that performs instantaneous traversal of all nodes using the recursive traversal code. If there is a run where the asynchronous observer misses nodes that the oracle observer observes, then it signifies a node (key) movement. If there is such a run, then our reasoner would recommend RCU framework. Otherwise, no additional reasoning is needed. An example of asynchronous and synchronous observer for linked list recursive traversal relation is shown Figure 4. The key-movement predicate definition is self-explanatory.

 \section{Overall Procedure and Soundness}
     Our reasoner performs the above four tasks based on a given data structure theory $\mathcal{T}$ and sequential data structure knowledge $\mathcal{K}$ and takes appropriate decisions on the structure of transformed concurrent code. It is also assumed that $\mathcal{K}$ contains the library of sequential data structure operations $\Sigma = \{\sigma_1, \sigma_2, .. \}$, where each $\sigma_i : S_{\mathcal{D}} \rightarrow S_{\mathcal{D}_{\bot}}$\footnote{$\bot$ signifies that $\sigma_i$ may not be applicable to all instances in $S_{\mathcal{D}}$} is mapping from one instance of data structure $\mathcal{D}$ to another. Without loss of generality we can assume $\Sigma = \{\sigma_1, \sigma_2\}$. We say that the operation $\sigma_i$ is applicable on an instance $\mathcal{D} \in S_{\mathcal{D}}$ if $pre(\sigma_1)$ is true in some model of $\mathcal{T}^\mathcal{R} \cup \mathcal{D}$. There exists a least $\delta \in S_{\mathcal{D}}$ such that each $\sigma_i \in \Sigma$ is applicable to $\delta$. This structure is assumed to be part of $\mathcal{T}^{\mathcal{R}}$. The instance $\delta$ is sufficient for the reasoning tasks performed in this paper. It is used in the soundness proof of the procedure later. The intuition behind choosing such an instance $\delta$ is that we need to model executions in which simultaneous operations contend to modify the data structure. If for some $\mathcal{D}'$ there is some $\sigma_i$ that is not applicable to $\mathcal{D}'$ then interference model $\mathcal{I}$ cannot model serialized concurrent execution faithfully, as there might be only a subset of operations modifying the data structure simultaneously. A safe concurrent algorithm must take into account interference effects from all destructive update operations in $\Sigma$. Few notations need their description, $pre(\sigma)$ denotes the precondition of some operation $\sigma$, $\mathit{falsify\_p}$ denotes the generated falsification predicate for fluent $p \in \mathcal{T}$, $Locks(\sigma)$ are the set of locks guessed according to some domain expert provided heuristic $\mathcal{H}$ on  $pre(\sigma)$, $Locks\_Adequate$ function checks the adequacy of guessed locks, $Program\_Order$ is the set of all valid program order permutations that preserve a given invariant $Inv(\bar{X})$ and finally, $KeyMove$ is the function that captures the presence of key-movement using similar predicates presented earlier. When the procedure recommends RCU for $\sigma$, then either key-movement is detected or an invariant is violated with any program order $\pi(\sigma)$. If the locks guessed by $\mathcal{H}$ are inadequate, then the user of our system can provide his/her own heuristic $\mathcal{H'}$ and retry.
      { 
     \begin{tabular}{p{8cm}}
    % \mathit{Fluents} =  \{p({\bar{X}})\ : p \in \mathcal{T} \land p \ is \ time \ dependent\} & \\  
    % \small
     $\mathit{Unfalsify} =  \{p(\bar{X}) : p \in \mathit{Fluents} \ \land \newline ~~~~~~~~~~~~~~~~~~~~~ (\mathcal{T}^\mathcal{R} \cup \mathcal{I} \cup \{\mathit{falsify\_p}\})
                                \models \neg p(\bar{X},T) \}$  \\
     $Locks_{\sigma} =  \{\mathcal{H}(pre(\sigma))\}$  \\
     $Locks\_Adequate(Locks_{\sigma})  = \newline ~~~~~~~~~~~~~~~~~~~~~~~ \left\{\begin{array}{lr}
                                          true, & \text{if } (\mathcal{I}^L \cup Locks_{\sigma}) \models \neg pre(\sigma) \\
                                          false, & \text{otherwise}
                                         \end{array}\right\}$  \\
     $Program\_Order(\sigma) =   \{\pi(\sigma) : (\mathcal{T}^{po} \cup \pi(\sigma) \cup Inv(\bar{X})) \newline 
                                      ~~~~~~~~~~~~~~~~~~~~~~~`~~~~~~~~~~~~~~~~~~~~~ \text{is satisfiable} \}$ \\
    $KeyMove(\sigma) = \newline \left\{\begin{array}{l}
                           \!\!\! false,  \text{if }  \\ 
                           \!\!\!       (\mathcal{T}^{\mathcal{R}} \cup 
                                      \mathcal{I} \cup \{trv, trv^{\mathcal{O}}, keymove_{\sigma}\}) 
                                     \models \neg keymove_{\sigma}  \\  
                           \!\!\! true  \ \text{otherwise}
                           \end{array}\!\!\!\right\}$
     \end{tabular}
     }
     \par     
     Without loss of generality assume that we are trying to transform 2 operations of some tree-based inductive data structure $\mathcal{D}$  their corresponding concurrent versions. Let $\sigma_1, \sigma_2$ denote the two operations. Again, without loss of generality that both $\sigma_1$ and $\sigma_2$ have a single basic block in their destructive update code. For External BSTs insert operation. there are four different pre-conditions and hence four basic blocks. But, as we show, the argument follows similarly if we consider single basic block. Because, $\mathcal{D}$ is inductive, we assume  that $\sigma_1$, $\sigma_2$ are applicable to countable infinite instances of in $S_{\mathcal{D}}$. Clearly, there exists a least instance $\delta \in S_{\mathcal{D}}$ such that $\sigma_1$ and $\sigma_2$ are applicable to $\delta$.  The agents (including interference) that perform $\sigma_1$ or $\sigma_2$ are always cautious with respect to their (permuted) sequential steps from $\mathcal{K}$. That is, after acquiring the desired locks, the agents post-check (validate) their respective preconditions $pre(\sigma_1)$ or $pre(\sigma_2)$ to ensure that the ``window" of modification is still intact and not modified in the time taken to acquire the locks. \\

% \section{Soundness of Procedure}  
 % \par
       
    \fbox{
 \begin{minipage}{0.97\linewidth}
    \begin{algorithmic}
          %  \small
 \Procedure{GenerateConcurrentCode}{$\sigma$}
   \State Code $\gets$ $\emptyset$
   \If{$Program\_Order(\sigma) = \emptyset$}
      \State {\textsc{RecommendRCU($\sigma$)}}
      \Return
   \EndIf
   \If{$KeyMove(\sigma) = true$}
             \State \textsc{RecommendRCU($\sigma$)}
              \Return
    \EndIf
   \If{$Locks\_Adequate(Locks\_{\sigma}) = \mathit{false}$}
        \State \textsc{RecommendRCU($\sigma$)}
              \Return
   \EndIf
        %\If{$KeyMove(\sigma) = \mathit{false}$}
        \State Code $\gets$ Code $\oplus$ \textsc{LockStmts($Locks_{\sigma}$)}%\footnote{$\oplus$ is simply textual code concatentation}
        %\EndIf
        \State Code $\gets$ Code $\oplus$ \textsc{Validate($pre(\sigma) \setminus \mathit{Unfalsify}$)}
        \State Code $\gets$ Code $\oplus$ $Program\_Order(\sigma)$
        \State Code $\gets$ Code $\oplus$ \textsc{UnlockStmts($Locks_{\sigma}$)}
 \EndProcedure
 \end{algorithmic}
 \end{minipage}
 }   
  \subsection{Soundness of Unfalsifiable Predicates} 
   \textit{Lemma 1:} If a time-dependent predicate $p(\bar{X}, T)$ (fluent) is unfalsifiable in $\mathcal{I}$ for some $\delta$, then it is unfalsifiable in any serialized concurrent execution of $\sigma_1$ and $\sigma_2$ \\ 
   \textit{Proof:}  $p(\bar{X})$ may belong to $pre(\sigma_1)$ or $pre(\sigma_2)$ (or both). We denote $\sigma_i$ to mean one of either $\sigma_1$ or $\sigma_2$. \\
   \textit{Case 1:}  $\sigma_i(\delta) = \delta'$ and every $\sigma_i$ is applicable to $\delta'$. Then we have no problem. 
    As conjuncts of $\sigma_1, \sigma_2$ are not falsified including $p(\bar{X})$. \\
   \textit{Case 2:}  $\sigma_i(\delta)$ = $\delta'$ and some $\sigma_j$ is not applicable to $\delta'$. If $p(\bar{X}) \in pre(\sigma_{j'}), j' \neq j$, then we have no problem. If otherwise, $p(\bar{X}) \in pre(\sigma_j)$, there must exist another predicate $p'(\bar{Y}) \in pre(\sigma_j)$ such that $p'(\bar{Y})$ is falsified. Otherwise, $\sigma_j$ would be applicable to $\delta'$ (as $p(\bar{X})$ is not falsified).
   From the above two cases, it is clear that any serialized run of operations $\sigma_1$ and $\sigma_2$ does not falsify $p(\bar{X})$.  
   \subsection{Lock Adequacy argument in $\mathcal{I}^{\mathcal{L}}$ is sound:}
   \textit{Lemma 2:} If the guessed locks for some $\sigma_j$ make $pre(\sigma_j)$ unfalsifiable in $\mathcal{I}^\mathcal{L}$, then $pre(\sigma_j)$ is unfalsifiable in any serialized execution of $\sigma_1$ and $\sigma_2$.
   \textit{Proof} similar to Lemma 1.

 \section{Experiments, Conclusion and Future Work}
     Our approach has been applied to Linked Lists, External BSTs and Internal BSTs (Table 1). Currently we are able to synthesize the concurrent versions of insert, delete for Linked Lists and External BSTs. Our reasoner can also recommend RCU framework for Internal BSTs due to key-movement missed by an asynchronous observer. \mycomment{The locks acquired on only the nodes that are modified in the sequential computation of Internal BST are inadequate. Clearly, for Internal BSTs delete operation the entire connected nodes in the window need to be locked. However, the well-formedness of Internal BST cannot be maintained while performing delete operations. Therefore, RCU is recommended.} Figure 5 shows an example for Linked Lists Insert.

\mycomment{
    \begin{table}
    {\scriptsize
     \begin{tabular}{|p{2cm}|p{1cm}|p{1cm}|p{1cm}|p{1cm}|}
       \hline
       Operations  & Task 1 & Task 2 & Task 3 & Task 4  \\
       \hline 
       Linked List Membership & N/A & N/A & No & N/A \\ 
       \hline
       Linked List Insert & Yes & Yes & No & No \\
       \hline
       Linked List Delete & Yes & Yes & No & No \\
       \hline
       External BST Membership & N/A & N/A & No & N/A \\ 
       \hline
       External BST Insert & ?? & Yes & No & No \\
       \hline
       Exteral BST Delete & ?? & Yes & No & No  \\
       \hline
       Internal BST Membership & N/A & N/A & Yes & N/A \\
       \hline
       Internal BST Insert & ?? & Yes & No & No \\
       \hline
       Internal BST Delete & ?? & No & Yes & No \\
       \hline
    
    \end{tabular}
    }
   \label{tab:results}
    \caption{Inferences made for each reasoning task for various data structures}
    \end{table}
    }

   % \section{Conclusion and Future Work}
        Our work presents the first step towards using commonsense reasoning to generate concurrent programs from sequential data structures knowledge. We have presented the challenges involved in the concurrent code generation and mechanized the reasoning tasks as performed by a human concurrency expert. The procedure described in this paper conforms to McCarthy's vision of building programs that have commonsense and manipulate formulas in first order logic \cite{mccarthy1960programs}.  Our future work aims to apply our technique to more data structures such as Red-Black Trees and AVL-Trees. In general, given the knowledge about a sequential data structure as well knowledge about the concept of concurrency, one should be able to generate suitable, correct versions of concurrent programs. We aim to generalize our technique to arbitrary data structures. Further, the only synchronization primitives we have addressed in this paper are \textit{locks}. However, there are more sophisticated atomic write instructions supported by modern multiprocessors such as \textit{Compare-and-Swap} \cite{valois1995lock}, \textit{Fetch-and-Add} \cite{heidelberger1990parallel}. They give rise to lock-free data structures. We plan to add knowledge about these primitives in our future work and generate more sophisticated concurrent programs.  
          \begin{table}
    {\scriptsize
     \begin{tabular}{|p{2cm}|p{1.5cm}|p{1cm}|p{1cm}|}
     \hline
      \textbf{Data Structures} &  \textbf{Membership} & \textbf{Insert} & \textbf{Delete} \\
      \hline
      Linked List &    No change &  Success &  Success \\
      \hline
      External BST & No change & Success & Success \\
      \hline
      Internal BST & No change & Success & RCU   \\
      \hline
     \end{tabular}
     }
  \label{tab:Results}
  \caption{Results of 4 reasoning tasks on few data structures}
  \end{table}

   \begin{figure}
      {%\fontsize{8}{9}\selectfont
       \small
       \begin{tabular}{p{8cm}} 
       {\tt 
       \fbox{
       \begin{tabular}{p{8cm}}
        \centering 
        
            $\{\underline{list}, reach(\fcolorbox{red}{white}{x}), edge(x,\fcolorbox{red}{white}{y}), \overline{\mathit{suffix(y)}}, k_x < k_{\tau} < k_y\}$ \newline
              \fbox{\fbox{x.next := $\tau$}}  \fcolorbox{green}{white}{$\langle1\rangle$}   \newline
            $\{not \ list, not \ reach(x), edge(x,y), edge(y,\tau),$ \newline
            $ \mathit{suffix(y), k_x < k_{\tau} < k_y}\}$ \newline
              \fbox{\fbox{$\tau$.next := y}}  \fcolorbox{green}{white}{$\langle2\rangle$} \newline   
           $\{\underline{list}, reach(x),  \mathit{suffix(y), k_x < k_{\tau} < k_y}$ \newline
            $reach(\tau), edge(x,\tau), edge(\tau,y)\}$ 
      \end{tabular}
      }
      } 
      { \tt
      \fbox{
    
      \begin{tabular}{p{8cm}}
      \centering
          $\{\underline{list}, pre_{ins}(x,\tau,y)\}$ \newline
             lock(\fcolorbox{red}{white}{x}) \newline               
             lock(\fcolorbox{red}{white}{y}) \newline
        %  $\{\underline{list}, pre\_ins?\footnote{has to validated post lock-acquisition}(x,\tau,y)\}$ \newline
            if \ validate($reach(x),edge(x,y), \overline{\textit{\st{suffix(y)}}}$, \newline
            $k_x < k_{\tau} < k_y)\{$  \newline 
          $\{\underline{list}, pre_{ins}(x,\tau,y)\}$ \newline
              \fbox{\fbox{$\tau$.next := y}} \fcolorbox{green}{white}{$\langle2\rangle$} \newline
          $\{\underline{list}, pre_{ins}(x,\tau,y)\}$ \newline
               \fbox{\fbox{x.next := $\tau$}} \fcolorbox{green}{white}{$\langle1\rangle$} \newline
          $\{\underline{list}, edge(x,\tau), edge(\tau, y), reach(\tau)\}$ \newline
          $\}\phantom{thequickbrownfoxjumpedoverthelazydogthequick}$
      \end{tabular}
      }
      }
      \end{tabular} \\
         \fcolorbox{red}{white}{\phantom{x}} : Correct Nodes identified as part of window (Task 2),  
         \fbox{\fbox{\phantom{lin}}} : Destructive update steps (numbered)  (Task 3),
         $\overline{su\!f\!\!f\!i\!x}$ : Unfalsifiable predicate $\mathit{suffix}$ (Task 1), 
         $\underline{list}$ : Concurrency Invariant (Task 3)
    \caption{Generated Fragment of Concurrent Code (bottom)}
    \label{fig:mapping}     
    }
    \end{figure}

        \bibliography{references}
\end{document}